# IMPROVING ACROSS-DATASET BRAIN TISSUE SEGMENTATION USING TRANSFORMER

**Vishwanatha M. Rao**[1]    **Zihan Wan**[2]    **Soroush Arabshahi**[1]    **David J. Ma**[1]    **Pin-Yu Lee**[1]

**Ye Tian**[1]    **Xuzhe Zhang**[1]    **Andrew F. Laine**[1]    **Jia Guo**[3,4]

January 21, 2022

## ABSTRACT

Brain tissue segmentation has demonstrated great utility in quantifying MRI data through Voxel-Based Morphometry and highlighting subtle structural changes associated with various conditions within the brain. However, manual segmentation is highly labor-intensive, and automated approaches have struggled due to properties inherent to MRI acquisition, leaving a great need for an effective segmentation tool. Despite the recent success of deep convolutional neural networks (CNNs) for brain tissue segmentation, many such solutions do not generalize well to new datasets, which is critical for a reliable solution. Transformers have demonstrated success in natural image segmentation and have recently been applied to 3D medical image segmentation tasks due to their ability to capture long-distance relationships in the input where the local receptive fields of CNNs struggle. This study introduces a novel CNN-Transformer hybrid architecture designed for brain tissue segmentation. We validate our model's performance across four multi-site T1w MRI datasets, covering different vendors, field strengths, scan parameters, time points, and neuropsychiatric conditions. In all situations, our model achieved the greatest generality and reliability. Our method is inherently robust and can serve as a valuable tool for brain-related T1w MRI studies. The code for the TABS network is available at: https://github.com/raovish6/TABS.

## 1 Introduction

Brain tissue segmentation represents an important application of medical image processing, in which an MRI image of the brain is segmented into three tissue types: gray matter (GM), white matter (WM), and cerebrospinal fluid (CSF). Brain tissue segmentation is a critical step in Voxel-Based Morphometry (VBM), a method used to quantitatively analyze MRI scans. VBM presents the ability to highlight subtle structural abnormalities by estimating differences in GM and WM brain tissue volume. As such, VBM has been prevalent for characterizing and monitoring conditions such as schizophrenia (Wright et al., 1995), Alzheimer's (Hirata et al., 2005), Huntingon's (Kassubek et al., 2004), and bipolar disorder (Nugent et al., 2006). VBM has also been used as an integral preprocessing tool in machine learning and deep learning based disease classification pipelines (Nemoto et al., 2021; Salvador et al., 2017). Outside of VBM, brain tissue segmentation is useful for characterizing tissue volume in particular regions of interest. It is often used with magnetic resonance spectroscopy to quantify metabolites by tissue type, and both techniques have been applied together to investigate morphological differences associated with various disorders (Auer et al., 2001; Bagory et al., 2011) as well as correct for metabolite measurements based on differing tissue fractions (Harris et al., 2015).

---

1: Department of Biomedical Engineering, Columbia University, New York, NY, USA

2: Department of Applied Mathematics, Columbia University, New York, NY, USA

3: Department of Psychiatry, Columbia University, New York, NY, USA

4: Mortimer B. Zuckerman Mind Brain Behavior Institute, Columbia University, New York, NY, USA

Despite the demonstrated utility of brain tissue segmentation, there is no universally accepted method capable of segmenting accurately and efficiently across a wide variety of datasets. Manual segmentation of brain tissue is extremely labor intensive, often impractical given larger datasets, and difficult even for experts. Alternatively, automated segmentation has proven challenging due to properties inherent to the MRI scans themselves. Changes in vendors or field strength have both been linked with increased variance in repeated scan measures (Han et al., 2006), and scans acquired through different imaging protocols tend to fluctuate more in terms of volumetric brain measures (Kruggel et al., 2010). Time of day, as well as time between scans, have been associated with variable tissue volume estimation (Karch et al., 2019) while neuropsychiatric conditions such as schizophrenia have been linked with subtle brain tissue anatomical changes (Koutsouleris et al., 2014). Together, these inconsistencies make it difficult for brain tissue segmentation solutions to be applicable across datasets of differing vendors, collection parameters, time points, and neuropsychiatric conditions.

Many of the earlier proposed automated solutions have depended on intensity thresholding (Dora et al., 2017), population-based atlases (Cabezas et al., 2011), clustering (Dora et al., 2017; Mahmood et al., 2015), statistical methods (Angelini et al., 2007; Greenspan et al., 2006; Marroquín et al., 2002; Zhang et al., 2000), and standard machine learning algorithms. Thresholding-based approaches often struggle to segment low contrast input images with overlapping brain tissue intensity histograms. Alternatively, atlas-based algorithm performance heavily depends on the quality of the population-derived brain atlas. While machine learning algorithms such as support vector machine (SVM) (Bauer et al., 2011), random forest (Dadar and Collins, 2021), and neural networks (Amiri et al., 2013) have demonstrated reasonable segmentation performance, their accuracy largely relies on the quality of manually extracted features. In general, many of these algorithms require *a priori* information to properly segment brain tissue, which is often not feasible to acquire for all new scans segmented. FSL FAST is a popular statistical brain tissue segmentation toolkit that combines Gaussian mixture models with hidden Markov random fields to achieve reliable segmentation performance across a variety of datasets (Zhang et al., 2000). However, segmentation via FAST is time consuming and therefore not ideal for many real-time segmentation applications.

Convolutional neural networks (CNNs) have recently emerged as a superior alternative to standard machine learning algorithms for classification-based brain segmentation given their feature-encoding capabilities (Akkus et al., 2017). CNNs have been found to outperform machine learning algorithms such as random forest and SVM specifically for brain tissue segmentation (Zhang et al., 2015). Following their introduction, many other CNN-based networks have been proposed for brain tissue segmentation (Khagi and Kwon, 2018; Moeskops et al., 2016) as well as brain tumor segmentation (Beers et al., 2017; Feng et al., 2020a; Mlynarski et al., 2019), including both 2D and 3D approaches. Unet represents one popular segmentation algorithm (Çiçek et al., 2016; Ronneberger et al., 2015), which consists of symmetric encoding and decoding convolutional operations that allows for the preservation of the initial image resolution following segmentation. Variants of Unet have been successfully applied to brain tissue segmentation achieving state-of-the-art performance. For example, one study achieved a DICE score of 0.988 using 3D Unet, which even outperformed human experts (Kolařík et al., 2018). More recently, 2D patch-based Unet and Unet-inspired implementations have gained traction (Lee et al., 2020; Yamanakkanavar and Lee, 2020) to better preserve and account for local details; such models have outperformed their non-patch-based variants.

Despite the impressive performance CNNs have demonstrated for brain tissue segmentation, they often struggle to generalize well when presented with new datasets. Many prior brain tissue segmentation approaches only report test performance on the same dataset upon which the model was trained. While such metrics validate the generality of the proposed model on MRI scans from the same dataset, they fail to quantify model performance across different datasets where changes in acquisition parameters can impact MRI image features and thus decrease the model's generality. Given the importance of brain tissue segmentation in VBM and pre-processing, it is not practical to retrain a CNN model every time a scan is obtained differently. As such, model generality is especially imperative to developing a widely applicable automated brain tissue segmentation solution.

Transformers are an alternative to CNNs that have recently demonstrated state-of-the-art results in natural image segmentation. Emerging evidence suggests that Transformers coupled with CNNs may improve performance and generalization for medical image segmentation tasks including brain tissue segmentation (Chen et al., 2021; Hatamizadeh et al., 2021; Sun et al., 2021; Wang et al., 2021). In this study, we sought to improve the traditional Unet architecture using Transformers to not only achieve higher brain tissue segmentation performance, but also generalize better across different datasets while remaining reliable. Here, we propose Transformer-based Automated Brain Tissue Segmentation (TABS), a new 3D CNN-Transformer hybrid deep learning architecture for brain tissue segmentation.

Our main contributions include:

1. A novel CNN-Transformer hybrid architecture designed for brain tissue segmentation.
2. We elucidate the benefits of embedding a Transformer module within a CNN encoder-decoder architecture for brain tissue segmentation.
3. After achieving improved within dataset performance, we are the first to rigorously demonstrate model generality and reliability across multiple vendors, field strengths, scan parameters, time points, and neuropsychiatric conditions.

## 2 Methods

### 2.1 Study Design

We conducted three experiments to evaluate model performance, generality, and reliability for brain tissue segmentation. The experimental pipeline for these experiments is visualized in **Figure 1**. First, we trained and tested all of the models on three separate datasets (DLBS, SALD, and IXI) of differing acquisition parameters along with an aggregate total dataset containing all of the scans combined. We then evaluated model generality across field strength and scanner parameters; models trained on 3T datasets were tested on the 1.5T dataset and models trained on 3T datasets were tested on one another. Finally, we extended our generalization testing to an alternate dataset (COBRE) containing test-retest repeated scans of both schizophrenia and healthy patients. We applied models pre-trained on the 3T SALD dataset to COBRE to give them the best chance of generalizing well, as SALD and COBRE were collected using similar acquisition parameters. We compared the reliability of TABS, the best generalizing model, to that of the ground truth by evaluating the similarity of outputs on the test-retest repeated scans. Given that each pair of scans were acquired from the same subject within a small time frame, we expected a more reliable tool to output very similar segmentation predictions across both scans.

We compared TABS to three other benchmark CNN models in our experiments: vanilla Unet, Unet-SE, and ResUnet. We chose Unet given its prior state of the art performance in 3D brain tissue segmentation (Kolařík et al., 2018), and we also compared to prior attempts at improving Unet including squeeze-excitation (SE) blocks (Hu et al., 2018) before each downsampling operation (Unet-SE) and residual connections (ResUnet) (Zhang et al., 2018). Moreover, given that the model architecture for TABS is identical to that of ResUnet except for the Vision Transformer, comparing to ResUnet allowed us to highlight the specific benefits conferred by the Transformer. All of the tested models were the same depth and encoded the same number of features. Finally, we also compared to FSL FAST, the tool used to generate the ground truths, in our reliability evaluation.

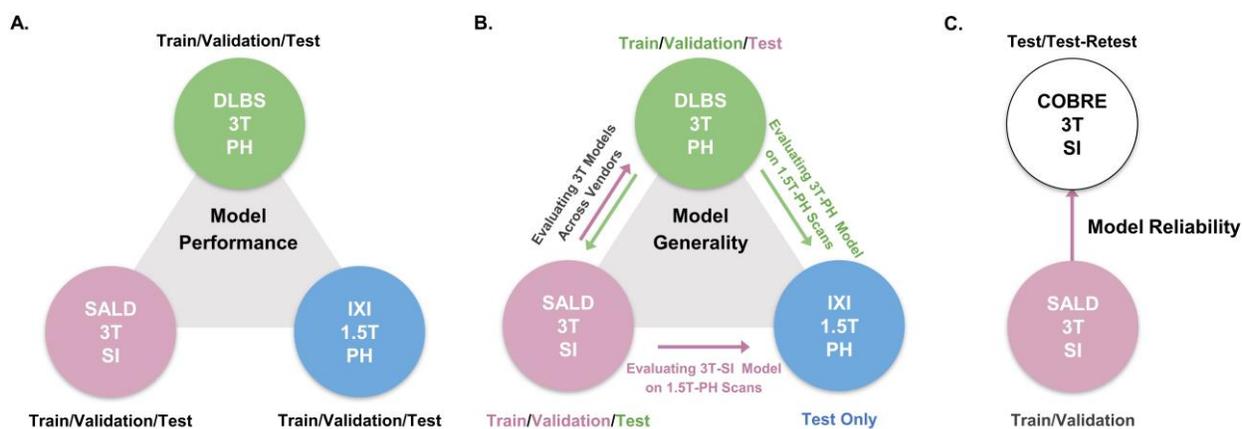

**Figure 1:** Overview of the experimental pipeline. **A.** Model performance test, where each model was trained and tested on individual datasets. **B.** Model generality test, where models pre-trained on 3T DLBS/SALD datasets were tested on one another and on the 1.5T IXI dataset. **C.** Model reliability test, where the best generalizing model to the COBRE dataset was compared to FAST based on similarity in segmentation outputs for repeated scans.

## 2.2 Data Selection and Pre-Processing

We collected MRI scans of healthy participants over a broad age range from three datasets for our first two experiments: DLBS (Rodrigue et al., 2012), SALD (Wei et al., 2018), and IXI (Biomedical Image Analysis Group, 2018). While they all use an MPRAGE sequence, the datasets vary in terms of their other acquisition parameters. Firstly, they differ by field strength, where DLBS and SALD contain 3T scans and IXI contains 1.5T scans. Moreover, all three datasets were acquired using different scanners, with the SALD dataset acquired using a Siemens manufactured scanner as opposed to Phillips. Lastly, the datasets differ in terms of scan parameters such as repetition/echo time and flip angle. We split each dataset into 3:1:1 train/validation/test groups while maintaining a broad age distribution across each subsection. The age distributions across these splits for each of these datasets are shown in **Figure 2a-c**. We also collected paired test-retest scans taken at different time points of healthy participants and schizophrenia patients from the COBRE dataset (Bustillo et al., 2017) for our third experiment. The demographic information and acquisition parameters for all four datasets are outlined in **Table 1**.

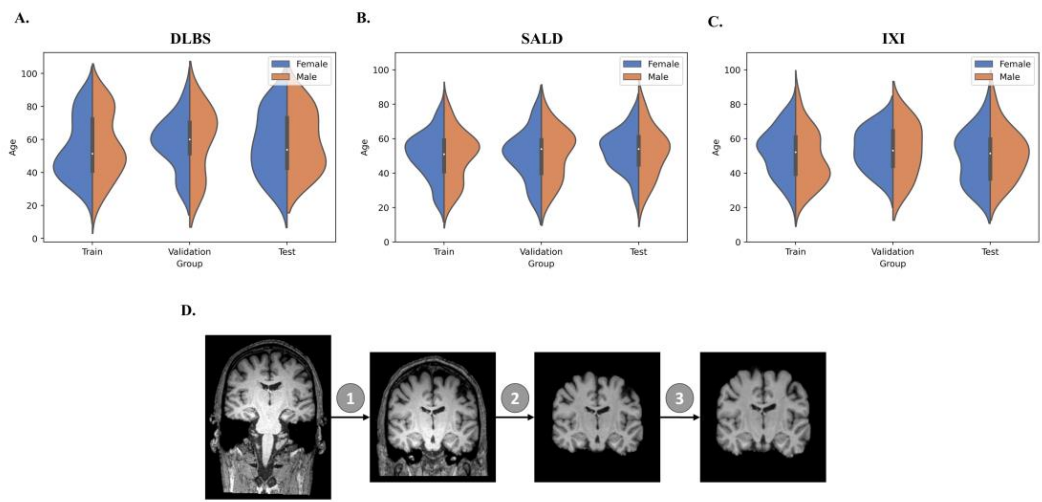

**Figure 2:** Data demographic and pre-processing visualization. **A-C.** Age distribution by gender for train/validation/test groups of DLBS, SALD, and IXI respectively. **D.** MRI preprocessing pipeline consisting of 1. Bias field correction 2. Brain extraction 3. Affine correction.

| Scan Parameters | DLBS | SALD | IXI | COBRE |
|---|---|---|---|---|
| Scanner | Philips Achieva | SIEMENS TrioTim | Phillips Intera | SIEMENS TrioTim |
| Field Strength | 3T | 3T | 1.5T | 3T |
| Sequence | MPRAGE | MPRAGE | MPRAGE | MPRAGE |
| Voxel Size (mm) | 1.0x1.0x1.0 | 1.0x1.0x1.0 | 1.0x1.0x1.0 | 1.0x1.0x1.0 |
| TR/TE (msec) | 8.10/3.70 | 1900/2.52 | 9.81/4.60 | 2530/1.64 |
| FA (degrees) | 12 | 90 | 8 | 7 |
| Number of Scans (Train/Validation/Test) | 129/43/43 | 170/56/57 | 137/45/46 | 0/0/358 (179 pairs) |
| Female % | 61.9 | 64.3 | 56.1 | 24.0 |
| Age Range (years) | 20-89 | 21-80 | 21-86 | 18-66 |
| Age Mean ± SD | 56.4 ± 18.2 | 50.6 ± 13.5 | 51.1 ± 14.2 | 38.3 ± 12.6 |

**Table 1:** MRI acquisition and demographic parameters for DLBS, SALD, IXI, and COBRE datasets.

We followed the initial pre-processing protocol outlined by Feng et al. (2020b) for all of the datasets, which includes bias field correction (Sled et al., 1998), brain extraction using FreeSurfer (Ségonne et al., 2004), and affine registration to the 1 mm$^3$ isotropic MNI152 brain template with trilinear interpolation using FSL FLIRT (Jenkinson et al. 2002). After these steps, the DLBS/SALD/IXI MRI images were 182x218x182, and the COBRE images were 193x229x193. We padded and cropped the images to reach an input dimension of 192x192x192, using a maximum intensity projection across all scans for each dataset to ensure that we did not remove important anatomical components. Finally, we normalized the intensities for each scan to values between -1 and 1. The pre-processing pipeline is shown in **Figure 2d**.

## 2.3 Model Architecture and Implementation

The architecture of our proposed model is shown in **Figure 3**. TABS is a ResUnet (Zhang et al., 2018) inspired model that consists of a 5-layered 3D CNN encoder and decoder. TABS takes an input dimension of 192x192x192, and the five encoder layers downsample the original image to $f$ x12x12x12, where $f$ represents the number of encoded features. For this specific implementation, we chose a $f$ value of 128. We follow the same "linear projection and learned positional embedding" operations introduced in Wang et al. (2021) to convert the encoded feature tensor into 512 tokenized vectors that are sequentially fed into the Transformer module in the order determined by the learned positional embeddings. Our Transformer encoder consists of 4 layers and 8 heads following the implementation initially described by Vaswani et al. (2017). The output of the Transformer is 512x1728, which we then reshape to 512x12x12x12 and reduce the feature dimensionality to $f$ via convolution. The decoder portion of the network reconstructs the image to the original input dimension, and a final convolution operation is applied to generate a 3-channel output with each channel corresponding to an individual tissue type. We used a Softmax activation function to ensure that the probabilities for each voxel across the three channels add up to 1.

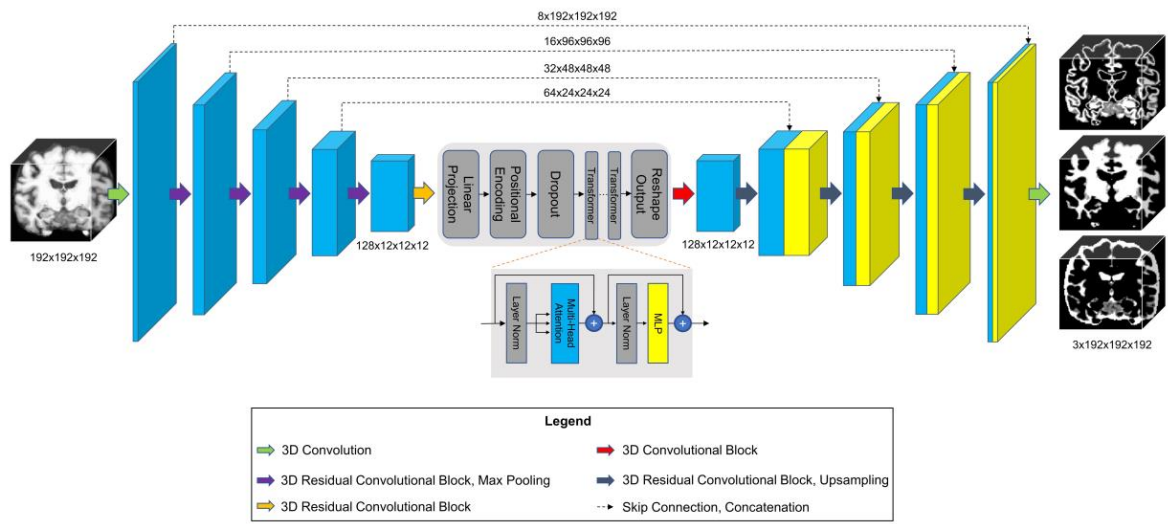

**Figure 3:** Model architecture for TABS, including a 5-layer encoder/decoder with a Vision Transformer between the encoder and decoder.

## 2.4 Training Protocol

All four models were trained using the same parameters described below. We trained for 350 epochs on the three individual datasets, while we trained over 200 epochs for the larger total dataset. We selected pre-trained models based on the best validation performance. We used FAST to generate ground truth probability maps for each brain tissue type and stacked and cropped them to generate a three-channel image matching the output shape of our models (3x192x192x192). The models were trained on three 24 GB NVIDIA Quadro 6000 graphical processing units using mean-squared-error (MSE) loss with a batch size of 3. We used group normalization as opposed to batch normalization due to group normalization's increased stability for smaller batch sizes (Wu and He, 2018). We trained using Adam (Kingma and Ba, 2014) as the optimization algorithm with a learning rate of 1E-5 and weight decay set to 1E-6.

## 2.5 Evaluation Metrics

All evaluation metrics were only taken for the portion of the outputs containing the brain, meaning that the background voxels outside of the segmentation field were not considered. Additionally, all metrics were calculated individually for each brain tissue type. Segmentation similarity using continuous probability estimates was quantified using Pearson correlation, Spearman correlation, and MSE. Segmentation maps for each tissue type were then generated from the probability estimations by taking the argmax along the channel axis. We generated binary maps for each tissue type based on the numerical value assigned to each voxel of the argmax output. Segmentation similarity between these binary maps was quantified using DICE Score, Jaccard Index, and Haussdorf Distance (HD) (Beauchemin et al., 1998). Performance was compared between models based on the higher absolute value of the metric.

# 3 Results

## 3.1 Model Performance

The performance results for each model trained and tested on DLBS, IXI, SALD, and Total datasets individually are reported in **Table 2**. TABS outperformed ResUnet, Unet-SE, and Unet on all the datasets for most metrics except for the 1.5T IXI dataset, where TABS outperformed Unet-SE and Unet while only performing slightly worse than ResUnet. TABS consistently achieves higher DICE/Jaccard metrics across all tissue types along with higher correlation and lower MSE on most tissue types. In general, all models performed better on WM and CSF as opposed to GM. **Figure 4** plots representative segmentation outputs for performance testing for each of the datasets.

| Project | Metrics | | TABS | | | ResUnet | | | Unet-SE | | | Unet | | |
|---|---|---|---|---|---|---|---|---|---|---|---|---|---|---|
| | | | Gray Matter | White Matter | CSF | Gray Matter | White Matter | CSF | Gray Matter | White Matter | CSF | Gray Matter | White Matter | CSF |
| DLBS | DICE | ↑ | **0.932 ± 0.024** | **0.954 ± 0.013** | **0.964 ± 0.010** | 0.929 ± 0.026 | 0.951 ± 0.013 | 0.963 ± 0.008 | 0.925 ± 0.026 | 0.951 ± 0.014 | 0.956 ± 0.009 | 0.924 ± 0.027 | 0.947 ± 0.014 | 0.959 ± 0.009 |
| | Jaccard Index | ↑ | **0.874 ± 0.041** | **0.913 ± 0.023** | **0.930 ± 0.018** | 0.868 ± 0.044 | 0.907 ± 0.023 | 0.928 ± 0.015 | 0.861 ± 0.045 | 0.907 ± 0.025 | 0.917 ± 0.017 | 0.859 ± 0.046 | 0.900 ± 0.025 | 0.922 ± 0.016 |
| | Pearson | ↑ | **0.965 ± 0.009** | **0.980 ± 0.006** | **0.984 ± 0.002** | 0.963 ± 0.010 | 0.979 ± 0.006 | 0.982 ± 0.002 | 0.961 ± 0.010 | 0.978 ± 0.007 | 0.979 ± 0.003 | 0.957 ± 0.012 | **0.980 ± 0.004** | 0.978 ± 0.002 |
| | Spearman | ↑ | **0.930 ± 0.014** | 0.868 ± 0.016 | **0.844 ± 0.013** | 0.928 ± 0.014 | 0.866 ± 0.014 | 0.838 ± 0.013 | 0.922 ± 0.013 | 0.868 ± 0.015 | 0.839 ± 0.012 | 0.921 ± 0.017 | **0.869 ± 0.016** | 0.825 ± 0.015 |
| | HD | ↓ | 9.179 ± 1.625 | 12.071 ± 2.142 | 10.795 ± 1.842 | 9.378 ± 1.547 | 12.107 ± 2.138 | 10.894 ± 2.077 | 7.788 ± 1.260 | 14.937 ± 2.444 | 11.763 ± 1.743 | **7.454 ± 1.184** | 14.278 ± 2.508 | 10.900 ± 1.841 |
| | MSE | ↓ | **0.012 ± 0.002** | **0.011 ± 0.003** | **0.007 ± 0.001** | 0.013 ± 0.002 | 0.013 ± 0.004 | 0.009 ± 0.001 | 0.018 ± 0.002 | 0.015 ± 0.004 | 0.016 ± 0.001 | 0.018 ± 0.003 | 0.027 ± 0.005 | 0.011 ± 0.001 |
| SALD | DICE | ↑ | **0.944 ± 0.017** | **0.959 ± 0.015** | **0.955 ± 0.014** | 0.941 ± 0.016 | 0.955 ± 0.013 | 0.954 ± 0.014 | 0.939 ± 0.018 | 0.956 ± 0.014 | 0.950 ± 0.016 | 0.939 ± 0.018 | 0.955 ± 0.015 | 0.950 ± 0.016 |
| | Jaccard Index | ↑ | **0.895 ± 0.030** | **0.922 ± 0.065** | **0.914 ± 0.026** | 0.888 ± 0.028 | 0.914 ± 0.024 | 0.912 ± 0.026 | 0.885 ± 0.031 | 0.915 ± 0.026 | 0.904 ± 0.028 | 0.886 ± 0.031 | 0.914 ± 0.026 | 0.906 ± 0.029 |
| | Pearson | ↑ | **0.969 ± 0.007** | **0.983 ± 0.006** | **0.982 ± 0.004** | 0.968 ± 0.007 | 0.980 ± 0.007 | 0.979 ± 0.005 | 0.964 ± 0.006 | 0.981 ± 0.007 | 0.978 ± 0.005 | 0.967 ± 0.007 | 0.980 ± 0.007 | 0.980 ± 0.005 |
| | Spearman | ↑ | **0.938 ± 0.007** | 0.864 ± 0.009 | 0.837 ± 0.015 | 0.938 ± 0.007 | 0.862 ± 0.009 | 0.832 ± 0.016 | 0.924 ± 0.010 | **0.864 ± 0.010** | 0.835 ± 0.015 | 0.937 ± 0.007 | 0.863 ± 0.009 | **0.837 ± 0.015** |
| | HD | ↓ | 7.489 ± 1.557 | 11.737 ± 1.834 | 11.255 ± 1.891 | 8.197 ± 1.568 | **11.012 ± 2.251** | **11.092 ± 1.894** | 7.386 ± 1.509 | 13.733 ± 2.685 | 11.712 ± 1.678 | **7.294 ± 1.553** | 13.171 ± 2.919 | 11.241 ± 1.866 |
| | MSE | ↓ | **0.011 ± 0.002** | **0.008 ± 0.003** | **0.007 ± 0.001** | 0.012 ± 0.002 | 0.010 ± 0.003 | 0.008 ± 0.001 | 0.014 ± 0.002 | 0.009 ± 0.003 | 0.011 ± 0.001 | 0.012 ± 0.002 | 0.010 ± 0.003 | 0.008 ± 0.001 |
| IXI | DICE | ↑ | **0.942 ± 0.020** | **0.958 ± 0.017** | **0.962 ± 0.010** | **0.943 ± 0.021** | **0.960 ± 0.017** | **0.962 ± 0.011** | 0.938 ± 0.019 | 0.958 ± 0.016 | 0.957 ± 0.012 | 0.939 ± 0.021 | 0.955 ± 0.018 | 0.960 ± 0.012 |
| | Jaccard Index | ↑ | **0.891 ± 0.034** | **0.920 ± 0.030** | **0.927 ± 0.018** | **0.892 ± 0.035** | **0.923 ± 0.029** | 0.926 ± 0.020 | 0.885 ± 0.032 | 0.919 ± 0.029 | 0.918 ± 0.021 | 0.885 ± 0.035 | 0.914 ± 0.010 | 0.924 ± 0.021 |
| | Pearson | ↑ | **0.969 ± 0.012** | **0.982 ± 0.009** | **0.984 ± 0.003** | **0.970 ± 0.011** | **0.984 ± 0.008** | **0.985 ± 0.003** | 0.961 ± 0.010 | 0.981 ± 0.009 | 0.980 ± 0.003 | 0.964 ± 0.012 | 0.982 ± 0.007 | 0.981 ± 0.002 |
| | Spearman | ↑ | **0.938 ± 0.012** | 0.848 ± 0.010 | 0.854 ± 0.014 | 0.937 ± 0.012 | **0.850 ± 0.009** | **0.857 ± 0.015** | 0.907 ± 0.015 | 0.848 ± 0.009 | 0.852 ± 0.014 | 0.937 ± 0.013 | **0.850 ± 0.009** | 0.845 ± 0.014 |
| | HD | ↓ | 8.785 ± 1.910 | 12.584 ± 2.361 | 10.715 ± 1.744 | 7.626 ± 2.042 | **11.993 ± 2.611** | **10.194 ± 1.729** | 7.479 ± 1.724 | 16.770 ± 3.034 | 11.426 ± 1.724 | **6.277 ± 1.040** | 14.159 ± 3.390 | 10.537 ± 1.722 |
| | MSE | ↓ | **0.011 ± 0.004** | 0.009 ± 0.004 | **0.007 ± 0.001** | **0.011 ± 0.003** | **0.008 ± 0.004** | **0.007 ± 0.001** | 0.016 ± 0.003 | **0.008 ± 0.004** | 0.013 ± 0.002 | 0.016 ± 0.003 | 0.020 ± 0.004 | 0.009 ± 0.001 |
| Total | DICE | ↑ | **0.945 ± 0.020** | **0.961 ± 0.014** | **0.963 ± 0.012** | 0.944 ± 0.019 | 0.960 ± 0.012 | **0.963 ± 0.013** | 0.941 ± 0.020 | 0.960 ± 0.013 | 0.959 ± 0.014 | 0.941 ± 0.022 | 0.959 ± 0.014 | 0.959 ± 0.015 |
| | Jaccard Index | ↑ | **0.896 ± 0.035** | **0.925 ± 0.026** | **0.929 ± 0.022** | 0.895 ± 0.033 | 0.924 ± 0.022 | **0.929 ± 0.023** | 0.889 ± 0.035 | 0.923 ± 0.024 | 0.921 ± 0.026 | 0.889 ± 0.037 | 0.921 ± 0.026 | 0.922 ± 0.027 |
| | Pearson | ↑ | **0.972 ± 0.009** | **0.984 ± 0.006** | **0.987 ± 0.003** | 0.971 ± 0.007 | **0.984 ± 0.005** | 0.985 ± 0.004 | 0.966 ± 0.009 | 0.983 ± 0.006 | 0.982 ± 0.004 | 0.970 ± 0.009 | 0.983 ± 0.006 | 0.984 ± 0.004 |
| | Spearman | ↑ | **0.938 ± 0.009** | 0.861 ± 0.014 | 0.849 ± 0.018 | 0.937 ± 0.010 | **0.862 ± 0.014** | 0.848 ± 0.018 | 0.922 ± 0.012 | 0.861 ± 0.014 | 0.846 ± 0.017 | 0.935 ± 0.011 | 0.861 ± 0.014 | 0.848 ± 0.019 |
| | HD | ↓ | 7.480 ± 1.824 | 12.273 ± 2.123 | 10.882 ± 1.7733 | 7.307 ± 1.800 | **11.714 ± 2.395** | **10.337 ± 1.814** | 7.114 ± 1.505 | 13.725 ± 3.041 | 11.078 ± 1.771 | **7.006 ± 1.506** | 12.423 ± 2.662 | 10.739 ± 1.816 |
| | MSE | ↓ | **0.010 ± 0.003** | **0.007 ± 0.003** | **0.005 ± 0.001** | **0.010 ± 0.002** | **0.007 ± 0.002** | 0.006 ± 0.001 | 0.014 ± 0.002 | 0.008 ± 0.003 | 0.010 ± 0.001 | 0.011 ± 0.002 | 0.008 ± 0.003 | 0.006 ± 0.001 |

**Table 2:** Performance results with each model trained and tested on individual datasets. Columns are ordered from left to right in descending model performance. Bold text indicates superior performance, with equally performing model metrics both bolded.

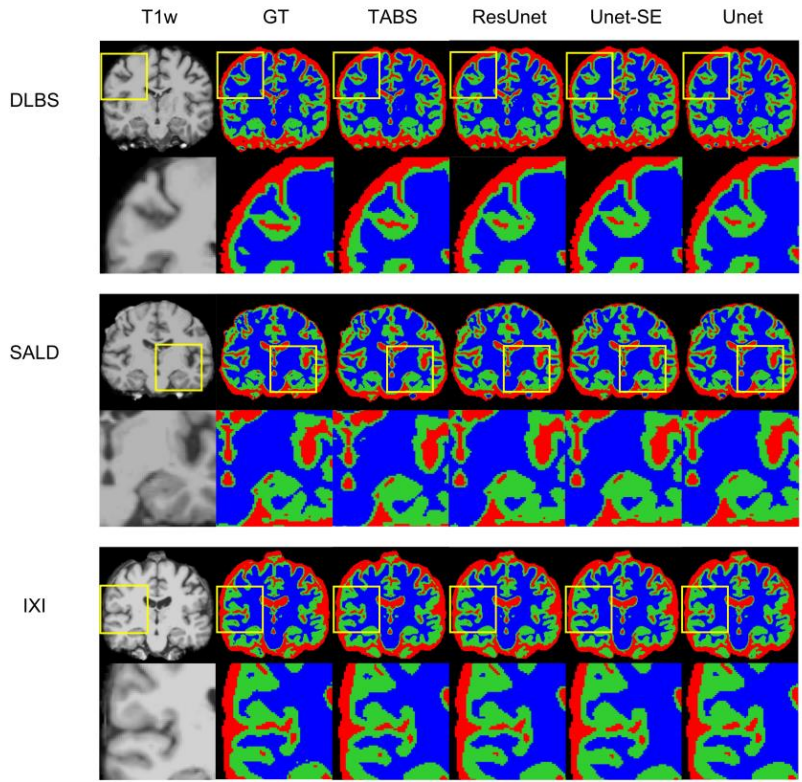

**Figure 4:** Visualization of model performance for DLBS, SALD, and IXI. Segmentation maps for the ground truth, TABS, and the 3 benchmark models are shown from left to right following the T1w scan. Zoom in regions are included below each image.

## 3.2 Model Generality – DLBS, IXI, and SALD

The generality results for all models trained on DLBS/SALD and applied to IXI as well as trained on DLBS/SALD and applied to SALD/DLBS are shown in **Table 3**. TABS generalized better across datasets on most metrics for the DLBS→IXI and SALD→DLBS tests, with higher DICE/Jaccard and correlation metrics for at least two tissue types. Additionally, for the SALD→IXI generalization test, TABS reached higher DICE/Jaccard metrics for both GM and WM. We observed that models trained on SALD performed better when applied to IXI than models trained on IXI itself. TABS also exhibited a similar increase in performance when pre-trained on DLBS and applied to IXI compared to TABS trained on IXI. Representative segmentation outputs for all models for each test scenario is shown in **Figure 5**.

| Project | Metrics | | TABS | | | Unet | | | Unet-SE | | | ResUnet | | |
|---|---|---|---|---|---|---|---|---|---|---|---|---|---|---|
| | | | Gray Matter | White Matter | CSF | Gray Matter | White Matter | CSF | Gray Matter | White Matter | CSF | Gray Matter | White Matter | CSF |
| DLBS → IXI | DICE | ↑ | **0.947 ± 0.021** | 0.964 ± 0.014 | **0.964 ± 0.010** | 0.943 ± 0.019 | **0.966 ± 0.011** | 0.953 ± 0.012 | 0.940 ± 0.020 | 0.964 ± 0.012 | 0.952 ± 0.013 | 0.938 ± 0.022 | 0.961 ± 0.011 | 0.955 ± 0.013 |
| | Jaccard Index | ↑ | **0.899 ± 0.036** | 0.931 ± 0.044 | **0.930 ± 0018** | 0.892 ± 0.033 | **0.935 ± 0.020** | 0.911 ± 0.023 | 0.887 ± 0.034 | 0.931 ± 0.022 | 0.908 ± 0.024 | 0.884 ± 0.038 | 0.925 ± 0.020 | 0.914 ± 0.023 |
| | Pearson | ↑ | **0.953 ± 0.018** | 0.978 ± 0.009 | **0.974 ± 0.007** | 0.937 ± 0.016 | 0.968 ± 0.007 | 0.958 ± 0.009 | 0.938 ± 0.020 | **0.978 ± 0.007** | 0.967 ± 0.008 | 0.937 ± 0.021 | 0.974 ± 0.008 | 0.965 ± 0.008 |
| | Spearman | ↑ | **0.923 ± 0.015** | 0.849 ± 0.010 | **0.825 ± 0.020** | 0.916 ± 0.011 | **0.849 ± 0.011** | 0.751 ± 0.033 | 0.901 ± 0.019 | **0.849 ± 0.010** | 0.817 ± 0.020 | 0.913 ± 0016 | **0.849 ± 0.010** | 0.780 ± 0.026 |
| | HD | ↓ | 8.416 ± 2.096 | 11.911 ± 2.384 | **11.437 ± 1.775** | **7.049 ± 1.455** | 13.218 ± 2.866 | 12.717 ± 1.786 | 7.484 ± 1.534 | 15.387 ± 2.957 | 13.099 ± 1.724 | 8.278 ± 1.866 | 11.334 ± 2.035 | 12.452 ± 1.721 |
| | MSE | ↓ | **0.016 ± 0.005** | 0.011 ± 0.004 | **0.011 ± 0.003** | 0.024 ± 0.004 | 0.025 ± 0.003 | 0.017 ± 0.003 | 0.025 ± 0.005 | 0.014 ± 0.003 | 0.019 ± 0.003 | 0.022 ± 0.006 | 0.014 ± 0.003 | 0.015 ± 0.003 |
| SALD → IXI | DICE | ↑ | **0.953 ± 0.019** | **0.970 ± 0.013** | 0.966 ± 0.010 | 0.950 ± 0.018 | 0.964 ± 0.015 | **0.968 ± 0.008** | 0.950 ± 0.018 | 0.967 ± 0.014 | 0.964 ± 0.018 | 0.949 ± 0.017 | 0.964 ± 0.013 | 0.967 ± 0.008 |
| | Jaccard Index | ↑ | **0.910 ± 0.033** | **0.941 ± 0.033** | 0.935 ± 0.019 | 0.905 ± 0.031 | 0.932 ± 0.026 | **0.937 ± 0.015** | 0.905 ± 0.031 | 0.937 ± 0.025 | 0.930 ± 0.015 | 0.903 ± 0.029 | 0.931 ± 0.023 | 0.935 ± 0.015 |
| | Pearson | ↑ | 0.958 ± 0.015 | 0.982 ± 0.007 | 0.978 ± 0.006 | **0.964 ± 0.010** | 0.982 ± 0.006 | **0.981 ± 0.004** | 0.957 ± 0.013 | **0.983 ± 0.007** | 0.978 ± 0.004 | 0.962 ± 0.011 | 0.982 ± 0.006 | 0.978 ± 0.005 |
| | Spearman | ↑ | 0.926 ± 0.014 | **0.851 ± 0.010** | 0.846 ± 0.012 | **0.931 ± 0.011** | 0.850 ± 0.010 | **0.851 ± 0.013** | 0.903 ± 0.019 | **0.851 ± 0.010** | 0.849 ± 0.012 | **0.931 ± 0.012** | 0.850 ± 0.010 | 0.843 ± 0.012 |
| | HD | ↓ | 9.152 ± 1.958 | 11.607 ± 2.205 | 11.074 ± 1.685 | 6.850 ± 1.536 | 17.200 ± 3.451 | 10.596 ± 1.726 | **6.540 ± 1.307** | 17.429 ± 3.215 | 11.341 ± 1.744 | 9.591 ± 1.667 | **11.549 ± 2.151** | 10.952 ± 1.650 |
| | MSE | ↓ | 0.014 ± 0.004 | **0.008 ± 0.003** | 0.009 ± 0.003 | **0.013 ± 0.003** | 0.009 ± 0.003 | **0.008 ± 0.003** | 0.016 ± 0.003 | **0.008 ± 0.003** | 0.012 ± 0.003 | **0.013 ± 0.003** | 0.009 ± 0.003 | 0.009 ± 0.003 |
| DLBS → SALD | DICE | ↑ | 0.931 ± 0.019 | 0.947 ± 0.015 | 0.944 ± 0.020 | **0.942 ± 0.014** | **0.958 ± 0.012** | **0.947 ± 0.012** | 0.937 ± 0.014 | 0.955 ± 0.011 | 0.944 ± 0.012 | 0.936 ± 0.015 | 0.956 ± 0.010 | 0.945 ± 0.014 |
| | Jaccard Index | ↑ | 0.871 ± 0.032 | 0.900 ± 0.067 | 0.894 ± 0.035 | **0.891 ± 0.024** | **0.920 ± 0.022** | **0.900 ± 0.022** | 0.882 ± 0.024 | 0.915 ± 0.020 | 0.894 ± 0.022 | 0.880 ± 0.026 | 0.915 ± 0.018 | 0.897 ± 0.025 |
| | Pearson | ↑ | **0.960 ± 0.012** | 0.976 ± 0.009 | **0.977 ± 0.007** | 0.951 ± 0.013 | 0.974 ± 0.006 | 0.959 ± 0.010 | 0.952 ± 0.013 | **0.978 ± 0.006** | 0.968 ± 0.008 | 0.949 ± 0.016 | 0.975 ± 0.007 | 0.966 ± 0.009 |
| | Spearman | ↑ | **0.933 ± 0.010** | 0.858 ± 0.008 | **0.828 ± 0.017** | 0.926 ± 0.010 | **0.861 ± 0.010** | 0.754 ± 0.036 | 0.920 ± 0.014 | **0.861 ± 0.009** | 0.813 ± 0.018 | 0.924 ± 0.012 | 0.860 ± 0.009 | 0.784 ± 0.027 |
| | HD | ↓ | 8.028 ± 1.603 | 12.200 ± 2.139 | **11.307 ± 1.897** | **7.817 ± 1.620** | 11.421 ± 2.367 | 13.463 ± 2.068 | 8.208 ± 1.516 | 13.226 ± 2.234 | 13.909 ± 1.939 | 8.631 ± 1.664 | **11.348 ± 1.946** | 13.158 ± 1.998 |
| | MSE | ↓ | **0.015 ± 0.004** | 0.024 ± 0.004 | **0.009 ± 0.002** | 0.021 ± 0.003 | 0.026 ± 0.003 | 0.015 ± 0.003 | 0.022 ± 0.003 | **0.014 ± 0.003** | 0.018 ± 0.002 | 0.019 ± 0.005 | **0.014 ± 0.003** | 0.013 ± 0.003 |
| SALD → DLBS | DICE | ↑ | **0.927 ± 0.040** | **0.953 ± 0.026** | 0.957 ± 0.013 | 0.905 ± 0.053 | 0.931 ± 0.038 | 0.950 ± 0.015 | 0.912 ± 0.047 | 0.937 ± 0.033 | 0.951 ± 0.016 | 0.921 ± 0.013 | 0.946 ± 0.026 | **0.957 ± 0.013** |
| | Jaccard Index | ↑ | **0.866 ± 0.064** | **0.912 ± 0.045** | 0.919 ± 0.023 | 0.830 ± 0.081 | 0.874 ± 0.064 | 0.905 ± 0.027 | 0.841 ± 0.073 | 0.883 ± 0.056 | 0.907 ± 0.028 | 0.855 ± 0.064 | 0.899 ± 0.045 | 0.917 ± 0.023 |
| | Pearson | ↑ | **0.952 ± 0.023** | **0.975 ± 0.015** | 0.976 ± 0.005 | 0.950 ± 0.037 | 0.967 ± 0.024 | **0.979 ± 0.004** | 0.951 ± 0.029 | 0.969 ± 0.022 | 0.976 ± 0.005 | **0.952 ± 0.023** | 0.973 ± 0.016 | 0.976 ± 0.004 |
| | Spearman | ↑ | 0.919 ± 0.020 | **0.861 ± 0.013** | 0.843 ± 0.014 | **0.920 ± 0.030** | 0.854 ± 0.014 | **0.848 ± 0.016** | 0.911 ± 0.021 | 0.856 ± 0.013 | 0.847 ± 0.005 | 0.919 ± 0.021 | 0.859 ± 0.014 | 0.840 ± 0.013 |
| | HD | ↓ | 10.330 ± 1.288 | 12.812 ± 1.892 | 11.174 ± 1.825 | 8.026 ± 1.293 | 15.906 ± 2.058 | 10.349 ± 1.920 | **7.781 ± 1.351** | 15.881 ± 2.167 | 10.498 ± 1.351 | 10.279 ± 1.232 | **12.142 ± 2.158** | 10.895 ± 1.671 |
| | MSE | ↓ | **0.017 ± 0.007** | **0.011 ± 0.007** | 0.010 ± 0.003 | **0.017 ± 0.010** | 0.015 ± 0.011 | **0.009 ± 0.002** | 0.018 ± 0.007 | 0.014 ± 0.010 | 0.013 ± 0.002 | **0.017 ± 0.007** | 0.013 ± 0.008 | 0.010 ± 0.002 |

**Table 3:** Generalization results across vendor, field strength, and scanning parameters. Models pre-trained on DLBS/SALD were applied to SALD/DLBS and IXI. Columns are ordered from left to right in descending model generality. Bold text indicates superior model performance, with equally performing model metrics both bolded.

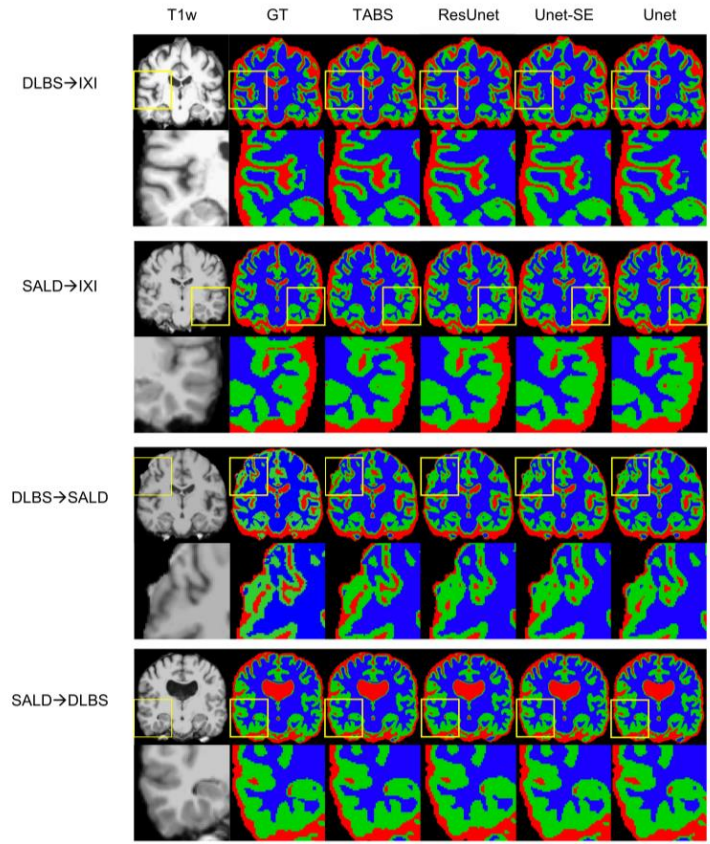

**Figure 5:** Visualization of model generality across vendors, field strength, and scanning parameters. Segmentation maps for the ground truth, TABS, and the 3 benchmark models are shown from left to right following the T1w scan. Zoom in regions are included below each image.

### 3.3 COBRE Generality

We extended our generalization testing to the COBRE dataset, consisting of healthy and schizophrenia test-retest repeated scans. The generalization performance for all models is reported in **Table 4**. TABS generalized better for GM and WM across the control, schizophrenia, and aggregate total dataset compared to the benchmark models for most metrics. Moreover, TABS also achieved higher DICE/Jaccard metrics for CSF for schizophrenia patients.

| Test | Metrics | | TABS | | | Unet | | | Unet-SE | | | ResUnet | | |
|---|---|---|---|---|---|---|---|---|---|---|---|---|---|---|
| | | | Gray Matter | White Matter | CSF | Gray Matter | White Matter | CSF | Gray Matter | White Matter | CSF | Gray Matter | White Matter | CSF |
| Control | DICE | ↑ | **0.872 ± 0.032** | **0.910 ± 0.024** | 0.902 ± 0.040 | 0.858 ± 0.016 | 0.894 ± 0.010 | 0.900 ± 0.025 | 0.846 ± 0.015 | 0.870 ± 0.012 | **0.909 ± 0.024** | 0.832 ± 0.016 | 0.853 ± 0.017 | 0.904 ± 0.024 |
| | Jaccard Index | ↑ | **0.774 ± 0.050** | **0.835 ± 0.042** | 0.824 ± 0.065 | 0.752 ± 0.025 | 0.809 ± 0.016 | 0.818 ± 0.041 | 0.733 ± 0.040 | 0.771 ± 0.018 | **0.833 ± 0.040** | 0.712 ± 0.023 | 0.744 ± 0.026 | 0.826 ± 0.040 |
| | Pearson | ↑ | **0.923 ± 0.025** | **0.952 ± 0.014** | **0.975 ± 0.006** | 0.918 ± 0.013 | 0.945 ± 0.007 | **0.975 ± 0.004** | 0.904 ± 0.016 | 0.923 ± 0.010 | 0.974 ± 0.005 | 0.882 ± 0.020 | 0.914 ± 0.014 | 0.973 ± 0.005 |
| | Spearman | ↑ | **0.902 ± 0.019** | 0.898 ± 0.010 | 0.754 ± 0.019 | 0.901 ± 0.011 | **0.901 ± 0.003** | 0.757 ± 0.020 | 0.888 ± 0.014 | 0.886 ± 0.007 | **0.759 ± 0.020** | 0.869 ± 0.018 | 0.855 ± 0.015 | 0.757 ± 0.020 |
| | HD | ↓ | 9.025 ± 1.559 | **9.672 ± 1.274** | 12.994 ± 2.524 | **8.488 ± 1.342** | 9.739 ± 1.209 | 13.619 ± 2.236 | 8.954 ± 1.430 | 10.400 ± 1.113 | 13.293 ± 2.378 | 11.023 ± 1.251 | 10.890 ± 1.087 | **12.797 ± 2.601** |
| | MSE | ↓ | **0.026 ± 0.008** | **0.025 ± 0.008** | **0.007 ± 0.002** | 0.028 ± 0.004 | 0.031 ± 0.004 | 0.008 ± 0.009 | 0.032 ± 0.005 | 0.040 ± 0.005 | 0.011 ± 0.001 | 0.042 ± 0.007 | 0.049 ± 0.008 | 0.009 ± 0.001 |
| Schiz | DICE | ↑ | **0.879 ± 0.033** | **0.919 ± 0.027** | **0.912 ± 0.040** | 0.850 ± 0.031 | 0.890 ± 0.012 | 0.903 ± 0.031 | 0.835 ± 0.018 | 0.865 ± 0.015 | 0.909 ± 0.029 | 0.820 ± 0.019 | 0.845 ± 0.019 | 0.907 ± 0.030 |
| | Jaccard Index | ↑ | **0.786 ± 0.053** | **0.851 ± 0.045** | **0.841 ± 0.065** | 0.739 ± 0.027 | 0.802 ± 0.019 | 0.824 ± 0.051 | 0.718 ± 0.026 | 0.762 ± 0.023 | 0.835 ± 0.048 | 0.695 ± 0.027 | 0.731 ± 0.028 | 0.831 ± 0.050 |
| | Pearson | ↑ | **0.929 ± 0.026** | **0.956 ± 0.016** | 0.974 ± 0.007 | 0.914 ± 0.016 | 0.942 ± 0.008 | **0.975 ± 0.005** | 0.896 ± 0.020 | 0.918 ± 0.013 | 0.974 ± 0.005 | 0.873 ± 0.024 | 0.907 ± 0.016 | 0.973 ± 0.005 |
| | Spearman | ↑ | **0.904 ± 0.019** | **0.902 ± 0.011** | 0.756 ± 0.022 | 0.896 ± 0.012 | 0.900 ± 0.004 | 0.763 ± 0.025 | 0.882 ± 0.017 | 0.883 ± 0.009 | **0.765 ± 0.026** | 0.861 ± 0.021 | 0.849 ± 0.016 | 0.763 ± 0.026 |
| | HD | ↓ | 8.809 ± 1.718 | **9.715 ± 1.386** | 13.013 ± 2.686 | **8.654 ± 1.520** | 10.074 ± 1.290 | 13.384 ± 2.497 | 9.051 ± 1.701 | 10.597 ± 1.202 | 13.079 ± 2.477 | 11.204 ± 1.285 | 11.334 ± 1.070 | **12.425 ± 2.807** |
| | MSE | ↓ | **0.024 ± 0.008** | **0.023 ± 0.008** | **0.008 ± 0.002** | 0.030 ± 0.005 | 0.033 ± 0.005 | **0.008 ± 0.001** | 0.035 ± 0.006 | 0.043 ± 0.007 | 0.011 ± 0.001 | 0.046 ± 0.008 | 0.054 ± 0.008 | 0.009 ± 0.001 |
| Total | DICE | ↑ | **0.875 ± 0.040** | **0.914 ± 0.026** | 0.907 ± 0.040 | 0.854 ± 0.018 | 0.892 ± 0.011 | 0.901 ± 0.028 | 0.841 ± 0.017 | 0.868 ± 0.013 | **0.909 ± 0.027** | 0.826 ± 0.018 | 0.849 ± 0.019 | 0.906 ± 0.027 |
| | Jaccard Index | ↑ | **0.780 ± 0.052** | **0.843 ± 0.044** | 0.832 ± 0.066 | 0.746 ± 0.027 | 0.806 ± 0.019 | 0.821 ± 0.046 | 0.725 ± 0.026 | 0.766 ± 0.021 | **0.834 ± 0.045** | 0.703 ± 0.027 | 0.738 ± 0.028 | 0.829 ± 0.045 |
| | Pearson | ↑ | **0.926 ± 0.025** | **0.954 ± 0.015** | 0.974 ± 0.007 | 0.916 ± 0.015 | 0.943 ± 0.008 | **0.975 ± 0.004** | 0.900 ± 0.019 | 0.920 ± 0.012 | 0.974 ± 0.005 | 0.878 ± 0.023 | 0.910 ± 0.015 | 0.973 ± 0.005 |
| | Spearman | ↑ | **0.903 ± 0.019** | 0.900 ± 0.011 | 0.755 ± 0.020 | 0.899 ± 0.023 | **0.901 ± 0.003** | 0.760 ± 0.023 | 0.885 ± 0.016 | 0.885 ± 0.008 | **0.762 ± 0.023** | 0.865 ± 0.020 | 0.852 ± 0.016 | 0.760 ± 0.023 |
| | HD | ↓ | 8.917 ± 1.642 | **9.694 ± 1.330** | 13.003 ± 2.603 | **8.571 ± 1.435** | 9.907 ± 1.260 | 13.501 ± 2.370 | 9.003 ± 1.571 | 10.499 ± 1.162 | 13.185 ± 2.427 | 11.114 ± 1.270 | 11.114 ± 1.100 | **12.610 ± 2.709** |
| | MSE | ↓ | **0.025 ± 0.008** | **0.024 ± 0.008** | **0.008 ± 0.002** | 0.029 ± 0.005 | 0.032 ± 0.005 | **0.008 ± 0.001** | 0.033 ± 0.006 | 0.042 ± 0.006 | 0.011 ± 0.001 | 0.044 ± 0.008 | 0.051 ± 0.009 | 0.009 ± 0.001 |

**Table 4:** Generalization results for each model pre-trained on SALD and applied to COBRE. Columns are ordered from left to right in descending model generality. Bold text indicates superior model performance, with equally performing model metrics both bolded.

### 3.4 COBRE Test-Retest

TABS showcased better reliability compared to FAST, the tool used to generate the ground truths. Similarity metrics between test-retest repeated images for both TABS and FAST are shown in **Table 5** for the control, schizophrenia, and total aggregate datasets. TABS proved consistently more reliable across almost all metrics for GM and CSF. Moreover, TABS reached a higher Pearson correlation and lower MSE over all tissue types, and only performed slightly worse than FAST on WM DICE/Jaccard. Representative segmentation outputs for paired repeated scans from both control and schizophrenia datasets are visualized in **Figure 6**.

| Test | Metrics | | TABS | | | FAST | | |
|---|---|---|---|---|---|---|---|---|
| | | | Gray Matter | White Matter | CSF | Gray Matter | White Matter | CSF |
| Control | DICE | ↑ | **0.959 ± 0.015** | **0.977 ± 0.008** | **0.954 ± 0.012** | 0.951 ± 0.015 | **0.977 ± 0.006** | 0.948 ± 0.012 |
| | Jaccard Index | ↑ | **0.922 ± 0.026** | 0.954 ± 0.016 | **0.912 ± 0.022** | 0.908 ± 0.026 | **0.955 ± 0.012** | 0.901 ± 0.021 |
| | Pearson | ↑ | **0.981 ± 0.010** | **0.994 ± 0.004** | **0.983 ± 0.007** | 0.968 ± 0.011 | 0.988 ± 0.005 | 0.867 ± 0.019 |
| | Spearman | ↑ | **0.980 ± 0.009** | 0.982 ± 0.009 | **0.973 ± 0.017** | 0.967 ± 0.011 | **0.985 ± 0.005** | 0.886 ± 0.018 |
| | HD | ↓ | **7.445 ± 1.513** | 8.066 ± 1.759 | 9.126 ± 1.942 | 7.489 ± 1.640 | **7.968 ± 1.507** | 10.624 ± 2.308 |
| | MSE | ↓ | **0.005 ± 0.003** | **0.002 ± 0.001** | **0.003 ± 0.002** | 0.011 ± 0.004 | 0.005 ± 0.002 | 0.030 ± 0.005 |
| Schiz | DICE | ↑ | **0.949 ± 0.020** | 0.972 ± 0.012 | **0.947 ± 0.016** | 0.941 ± 0.022 | **0.974 ± 0.009** | 0.942 ± 0.015 |
| | Jaccard Index | ↑ | **0.904 ± 0.036** | 0.946 ± 0.021 | **0.899 ± 0.028** | 0.890 ± 0.038 | **0.949 ± 0.017** | 0.891 ± 0.026 |
| | Pearson | ↑ | **0.974 ± 0.016** | **0.992 ± 0.007** | **0.978 ± 0.010** | 0.961 ± 0.018 | 0.985 ± 0.008 | 0.856 ± 0.024 |
| | Spearman | ↑ | **0.973 ± 0.014** | 0.978 ± 0.013 | **0.962 ± 0.028** | 0.959 ± 0.017 | **0.982 ± 0.008** | 0.875 ± 0.023 |
| | HD | ↓ | 7.779 ± 1.453 | 8.078 ± 1.674 | 10.068 ± 2.551 | **7.552 ± 1.531** | **7.990 ± 1.542** | 10.424 ± 2.445 |
| | MSE | ↓ | **0.007 ± 0.005** | **0.003 ± 0.002** | **0.004 ± 0.002** | 0.013 ± 0.006 | 0.006 ± 0.003 | 0.033 ± 0.007 |
| Total | DICE | ↑ | **0.954 ± 0.018** | 0.974 ± 0.010 | **0.950 ± 0.015** | 0.946 ± 0.019 | **0.975 ± 0.008** | 0.945 ± 0.014 |
| | Jaccard Index | ↑ | **0.913 ± 0.033** | 0.950 ± 0.019 | **0.906 ± 0.026** | 0.899 ± 0.034 | **0.952 ± 0.015** | 0.896 ± 0.024 |
| | Pearson | ↑ | **0.978 ± 0.014** | **0.993 ± 0.006** | **0.981 ± 0.009** | 0.964 ± 0.016 | 0.986 ± 0.007 | 0.861 ± 0.022 |
| | Spearman | ↑ | **0.976 ± 0.013** | 0.980 ± 0.011 | **0.967 ± 0.024** | 0.963 ± 0.015 | **0.983 ± 0.007** | 0.880 ± 0.022 |
| | HD | ↓ | 7.613 ± 1.488 | 8.072 ± 1.712 | 9.600 ± 2.311 | **7.506 ± 1.582** | **7.979 ± 1.521** | 10.524 ± 2.374 |
| | MSE | ↓ | **0.006 ± 0.004** | **0.002 ± 0.002** | **0.004 ± 0.002** | 0.012 ± 0.005 | 0.006 ± 0.003 | 0.032 ± 0.007 |

**Table 5:** Test-retest reliability results across time points and neuropsychiatric condition for TABS compared to FAST (ground truth) for control, schizophrenia, and aggregate total datasets from COBRE. Bold text indicates superior model performance, with equally performing model metrics both bolded.

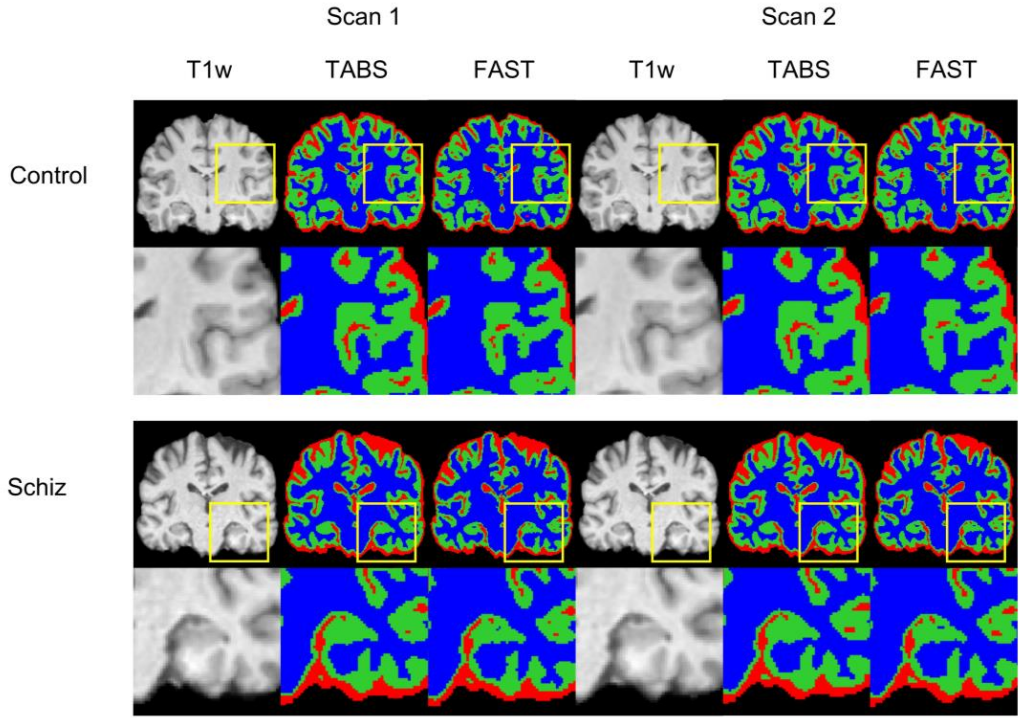

**Figure 6:** Visualization of test-retest reliability results across time points and neuropsychiatric conditions. Segmentation maps for TABS and FAST following the T1w scan are shown for each pair of repeated scans for control and schizophrenia groups. Zoom in regions are included below each image.

## 4 Discussion

In this study, we present TABS, a new Transformer-CNN hybrid deep learning architecture designed for brain tissue segmentation. TABS showcased superior performance compared to prior state-of-the-art CNN implementations while also generalizing exceptionally well across datasets and remaining reliable between paired test-retest scans. These traits are critical to developing a useful and more widely applicable brain tissue segmentation toolkit. Through TABS, we also demonstrate the methodological utility using a Vision Transformer to improve the Unet architecture for brain tissue segmentation.

Our experimental protocol was designed to elucidate the real-world applicability of TABS compared to various benchmark models. The datasets included in this study were chosen with the goal of emulating the extreme differences in MRI input a brain tissue segmentation algorithm would receive in real-world applications; the DLBS, SALD, and IXI datasets varied in terms of manufacturer, field strengths, and scanner parameters. Moreover, our test-retest dataset consisted of repeated scans from schizophrenia and healthy patients taken at different time points, presenting an even more challenging segmentation task. Due to these factors, we believe our evaluation methodology accurately captures the versatility of TABS.

### 4.1 Superior Performance for TABS Compared to Benchmark Unet Models

We found that TABS was the best performing model when trained and tested on the same dataset. While TABS achieved significantly higher performance than both Unet and Unet-SE, we observed marginal performance benefits over ResUnet. We hypothesize that the residual connections are responsible for the bulk of the performance gain over the traditional Unet models, with the Transformer module providing a small but consistent performance increase within the datasets.

### 4.2 Superior Generality for TABS Compared to Benchmark Unet Models

TABS generalized the best on most datasets compared to the benchmark Unet models. The most significant generalization differences we observed were between TABS and ResUnet. Given that their model architectures are identical except for the Transformer, we believe that the addition of the Transformer significantly improves model generality. CNNs are not well suited to capture long-range dependencies in the input image due to the local receptive fields of convolutional kernels. We believe that this property could make Transformer-based networks agnostic to dataset-specific variations and thus more generalizable. The addition of the Transformer allows TABS to preserve and even improve the within dataset performance conferred by residual connections while also generalizing better than the vanilla Unet, where ResUnet struggled.

We also noticed that all of the models tested improved in performance when trained on SALD and applied to IXI as opposed to training on IXI itself. This disparity could be due to the difference in field strength: the higher quality 3T MRI images from SALD may provide more globally relevant features than the 1.5T MRI images from IXI. However, for TABS specifically, we observed this same effect when pre-trained on 3T DLBS scans. These results indicate that TABS can potentially take better advantage of higher quality training data compared to the benchmark models.

Finally, we found that TABS generalized the best on an alternate COBRE dataset consisting of both healthy and schizophrenia scans. Schizophrenia patients often reflect subtle anatomical differences compared to healthy subjects, such as alterations in GM volume (Koutsouleris et al., 2014). These changes make generalizing to the schizophrenia dataset an especially difficult task. Additionally, the mean age of the COBRE dataset was slightly lower than the datasets TABS was originally trained on, making generalizing to COBRE potentially even more challenging. TABS generalized the best compared to the benchmark models on the overall COBRE dataset, with even more pronounced differences for the schizophrenia portion. Therefore, we believe that TABS may excel in more difficult segmentation cases where standard Unet models yield errors.

### 4.3 Superior Reliability for TABS Compared to FAST

Finally, our test-retest experiment highlights the reliability of TABS, the best generalizing model on the COBRE dataset, compared with the ground truth FAST. The test-retest repeated scans used in this study were taken from the same patient within a short time frame, meaning that we expected minimal differences in the segmentation output. One of the primary advantages of FAST has been its generality and reliability. Through this test, we find that TABS not only generalizes well on the COBRE dataset, but also maintains this performance more reliably than FAST.

### 4.4 Limitations and Future work

In general, 3D CNN models require a large amount of computational power to efficiently train. While we were able to use full resolution MRI inputs for our model, we were limited to a batch size of 3 due to memory constraints. Using a larger batch size may have resulted in better performance. Additionally, even though we trained TABS on three large datasets, our performance could be further improved by increasing our sample size. Recent findings suggest that patch-based 2D CNN approaches perform better than non-patch-based variants for brain tissue segmentation (Lee et al., 2020; Yamanakkanavar et al., 2020). As such, we believe that we could extend TABS to a patch-based 3D model in future studies to better capture local information that may be lost by processing the entire image at once.

## 5 Conclusion

In conclusion, we believe TABS represents a compelling brain tissue segmentation alternative. TABS performs and generalizes better than comparable state-of-the-art CNN models across vendor, field strength, scan parameters, and neuropsychiatric condition while remaining consistent across time points. Our results also demonstrate that the embedding of a Transformer module between the encoder and decoder portions of a CNN architecture represents an efficient method to improve brain tissue segmentation performance and generality.

### Data and Code Availability Statement

The code used in this project is proprietary. The code for the TABS model is available at https://github.com/raovish6/TABS, and the entire TABS package is available upon request of the corresponding author. The code for TABS is © 2021 The Trustees of Columbia University in the City of New York. This work may be reproduced and distributed for academic non-commercial purposes only.

## References

Auer, D.P., Wilke, M., Grabner, A., Heidenreich, J.O., Bronisch, T. and Wetter, T.C., 2001. Reduced NAA in the thalamus and altered membrane and glial metabolism in schizophrenic patients detected by 1H-MRS and tissue segmentation. *Schizophrenia research*, 52(1-2), pp.87-99.

Angelini, E.D., Song, T., Mensh, B.D. and Laine, A.F., 2007. Brain MRI segmentation with multiphase minimal partitioning: A comparative study. *International Journal of Biomedical Imaging*, *2007*.

Amiri, S., Movahedi, M.M., Kazemi, K. and Parsaei, H., 2013. An automated MR image segmentation system using multi-layer perceptron neural network. *Journal of biomedical physics & engineering*, *3*(4), p.115.

Akkus, Z., Galimzianova, A., Hoogi, A., Rubin, D.L. and Erickson, B.J., 2017. Deep learning for brain MRI segmentation: state of the art and future directions. *Journal of digital imaging*, *30*(4), pp.449-459.

Beauchemin, M., Thomson, K.P. and Edwards, G., 1998. On the Hausdorff distance used for the evaluation of segmentation results. *Canadian journal of remote sensing*, *24*(1), pp.3-8.

Bagory, M., Durand-Dubief, F., Ibarrola, D., Comte, J.C., Cotton, F., Confavreux, C. and Sappey-Marinier, D., 2011. Implementation of an Absolute Brain $^1$H-MRS Quantification Method to Assess Different Tissue Alterations in Multiple Sclerosis. *IEEE transactions on biomedical engineering*, 59(10), pp.2687-2694.

Bauer, S., Nolte, L.P. and Reyes, M., 2011, September. Fully automatic segmentation of brain tumor images using support vector machine classification in combination with hierarchical conditional random field regularization. In *international conference on medical image computing and computer-assisted intervention* (pp. 354-361). Springer, Berlin, Heidelberg.

Beers, A., Chang, K., Brown, J., Sartor, E., Mammen, C.P., Gerstner, E., Rosen, B. and Kalpathy-Cramer, J., 2017. Sequential 3d u-nets for biologically-informed brain tumor segmentation. *arXiv preprint arXiv:1709.02967*.

[dataset] Bustillo, J.R., Jones, T., Chen, H., Lemke, N., Abbott, C., Qualls, C., Stromberg, S., Canive, J. and Gasparovic, C., 2017. Glutamatergic and neuronal dysfunction in gray and white matter: a spectroscopic imaging study in a large schizophrenia sample. *Schizophrenia bulletin*, *43*(3), pp.611-619.

[dataset] Biomedical Image Analysis Group, Imperial College London, and Centre for the Developing Brain, King's College London, 2018. Information eXtraction from Images. https://brain-development.org/ixi-dataset/.

Cabezas, M., Oliver, A., Lladó, X., Freixenet, J. and Cuadra, M.B., 2011. A review of atlas-based segmentation for magnetic resonance brain images. *Computer methods and programs in biomedicine*, *104*(3), pp.e158-e177.

Çiçek, Ö., Abdulkadir, A., Lienkamp, S.S., Brox, T. and Ronneberger, O., 2016, October. 3D U-Net: learning dense volumetric segmentation from sparse annotation. In *International conference on medical image computing and computer-assisted intervention* (pp. 424-432). Springer, Cham.

Chen, J., Lu, Y., Yu, Q., Luo, X., Adeli, E., Wang, Y., Lu, L., Yuille, A.L. and Zhou, Y., 2021. Transunet: Transformers make strong encoders for medical image segmentation. *arXiv preprint arXiv:2102.04306*.

Dora, L., Agrawal, S., Panda, R. and Abraham, A., 2017. State-of-the-art methods for brain tissue segmentation: A review. *IEEE reviews in biomedical engineering*, *10*, pp.235-249.

Dadar, M. and Collins, D.L., 2021. BISON: Brain tissue segmentation pipeline using T1-weighted magnetic resonance images and a random forest classifier. *Magnetic Resonance in Medicine*, *85*(4), pp.1881-1894.

Feng, X., Lipton, Z.C., Yang, J., Small, S.A., Provenzano, F.A., Alzheimer's Disease Neuroimaging Initiative and Frontotemporal Lobar Degeneration Neuroimaging Initiative, 2020. Estimating brain age based on a uniform healthy population with deep learning and structural magnetic resonance imaging. *Neurobiology of aging*, *91*, pp.15-25.